\documentclass{article}
\usepackage{amssymb}

\usepackage{graphicx}
\usepackage{amsmath}

\input{tcilatex}

\begin{document}

\title{Preferred frame and two meanings of time:\\
diagonal form of the \textit{Lorentz-boost} transformation matrix}
\author{Paul\ Korbel \\
%EndAName
Institute of Physics, Technical University in Cracow, \\
ulica Podcharzych 1, 30-084 Cracow, Poland}
\maketitle

\begin{abstract}
The main purpose of this paper is to rethink the relativity issue within the
framework of the fundamental postulates of quantum mechanics. The aspect of
so-called ``double special relativity'' (DSR) is a starting point in our
discussion. The three elementary ideas were involved to show that special
relativity may be treated as an integral part of quantum mechanics. These
ideas (or observations) are: (1) the necessity of distinguishing the two
time meanings, namely: (i) the \textit{vital }one\textit{,} referring to
description of system evolution, and (ii)\ the \textit{frozen }one,
referring to energy measure by means of inverse time units; (2) the
existence of the energy-momentum (and time-distance) comparison scale in
relativistic description; and (3) a possibility of introduction of mass by
means of a light-cone frame description. The resulting quantum-mechanical
analysis allows us to find diagonal form of the \textit{Lorentz-boost}
transformation matrix and thus to relate the interval invariant relativity
principle with the principles of quantum mechanics. The manner, in which the
diagonal form of the transformation matrix was found, shows that covariant
description itself is a preferred frame description and that the time that
undergoes relativistic transformation rules is the \textit{frozen time},
whereas the \textit{vital time} is the Lorentz invariant. A generalized form
of the Heisenberg uncertainty principle, proposed by Witten \cite{witten},
is derived. It turns out, that this form is equivalent to the one know from
the analysis of the covariant harmonic oscillator given by Kim and Noz \cite
{Kbook}. As a by-product of this analysis one finds that special relativity
itself preserves the Planck length, however, a particle cannot be seen any
longer as a material point, but rather as an extended quantum object.
\end{abstract}

\newpage

\section{Introduction}

There is no doubt that the special theory of relativity gained the great
success in the field of high energy physics, where the local description
performed in momentum space dominates. In fact, high energy physics
experiment may be ``placed'' in a middle of energy scale, i.e. much above
the energy range proper for the atomic physics and much below the Planck
energy, playing the role of fundamental constant in quantum gravity.
However, for some time now, it seems that traditional relativity formulation
needs a refinement. Some physical examples taken from the ``both ends'' of
the sketched energy scale, seem to confirm such thesis.

In low energy limit we deal with the wide class of, so-called,
gedankenexperiments (originated from the well-known Einstein, Podolsky and
Rosen paradox \cite{epr}), which, by breaking the Bell inequalities \cite
{bellin}, manifest a strong nonlocality in quantum mechanics. A very
spectacular effect of two-photon entanglement was observed on a distance
exceeding 10 km \cite{10km}. Another aspect of gedankenexperiment is
non-causal information transfer between distant parts of the system \cite
{nonc}. Such behavior is blatantly in contradiction with Einstein locality
principle \cite{eloc}, according to which ``if $S_{1}$ and $S_{2}$ are two
systems that have interacted in the past but now are arbitrarily distant,
the real, factual situation of system $S_{1}$ does not depend on what is
done with system $S_{2}$, which is spatially separated from the former.''

An example of another obstacle resulting from orthodox relativity approach
concerns a number of problems in cosmology and quantum gravity with
fundamental question about the meaning of the Planck length $l_{p}=\sqrt{%
hG/\left( 2\pi c^{3}\right) }\sim 1.6\cdot 10^{-35}m$ (being the combination
of the Planck constant $h$, the gravitational constant $G$ and the
speed-of-light $c$) or its inverse, the Planck energy $E_{p}=1/l_{p}\sim
10^{19}GeV$. For many authors such energy scale acts as a threshold between
the known and unknown physical phenomena (e.g.\cite{grcoex}). Therefore,
constructing a new theory one expects the traditional relativity approach to
break down beyond the threshold but not beneath, where the gravitational
field may be weak or even absent. Then, the question asked by Magueijo and
Smolin \cite{MS} was ``in whose reference frame are $l_{p}$ and $E_{p}$ the
thresholds for new phenomena'', because, due to the effect of
Lorentz-Fitzgerald contraction the answer is not clear. On the other hand,
``it would be paradoxical if different observers disagreed on whether
quantum space-time effects are present in a process'', as stated by
Amelino-Camelia \cite{AC}. The class of recently investigated models known
as ``deformed'' or ``doubly special relativity'' (DSR) \cite{rev} is
intended to overcome this difficulty.

Actually, the idea of changing the postulates of special relativity by
introduction of the second absolute length scale already was introduced by
T. G. Pavlopoulos \cite{Pavlo}, in order to remove divergencies present in
relativistic field equations. This idea, next, was rediscovered by
Amelino-Camelia \cite{AC2} where the Planck length scale, another one
alongside the speed of light, was introduced, in hope to eliminate a
conflict between the assumed fundamental role of the Planck length in the
structure of space-time and the Lorentz-Fitzgerald contraction. Thus,
proposed name for the new theory was ``double special relativity'' \cite{AC3}%
. The introduced absolute length scale has set the maximum momentum value
(however, with energy has been left unbounded).

In the approach given in \cite{MS} the absolute energy and/or length scale
was intrinsically built-in into special relativity and set both the maximum
energy and momentum. The construction was based on modified action of the
Lorentz group acting on momentum space. This action, in general, is
nonlinear but reduces to the usual linear action in low energy limit.
However, the introduction of the absolute energy scale confronts the
complete relativity of inertial frames. On the other hand, the maintenance
of complete relativity is one of the major purpose of DSR models \cite
{AC3,MS}, what induces the search for general argument for relativity
modification other then the existence of a preferred frame. One of the claim
is that DSR models should emerge from a fundamental quantum gravity theory 
\cite{AC4}. Another possibility was created by so-called varying speed of
light (VSL) theories. For example, the VSL\ model discussed in \cite{rev2},
with a frequency dependent speed of light provided a basis for invariant
energy (or length) scale description. The both approaches, however, are
burden with difficulties the source of which in large extent is the same,
namely, non-linear relativity formulation (also) in position space \cite{KMM}%
. To simplify, in ``gravity approach'' one meets a problem of metric
definition, where, in the case of non-liner relativity realization at high
energies (or low distances), the concept of metric disintegrates \cite{MS}.
The VSL model, in turn, involves a problem of consistent velocity
determination simultaneously in momentum and in position space. Another
complication resulting from the loss of linearity provide descriptions of
many-particle systems, for which the kinematic relations of single particles
are not valid. So, the difficulties related to fixing the problems generated
by non-linear relativity, one may say, exceed somehow the very problem of
paradox being the source of DSR\ approach. The numerous papers cited in \cite
{rev2} gives voice to this.

A formulation based on a preferred frame concept may be an alternative to
DSR theory. As suggested in \cite{AM} a serious candidate for the preferred
frame is the cosmological frame. However, most physicists do not like
preferred frames. Why? In opinion of Magueijo \cite{rev2} ``this is more due
to mathematical or esthetical reasons then anything else: covariance and
background independence have been regarded as highly cherished mathematical
assets since the proposal of general gravity''. But the other possible
answer might be that, so far, the concept of preferred frame is simply
misunderstood and, in fact, there is no need to refer to the one
particularly chosen physical frame to make use of the \textit{preferred
frame description. }Such point of view, of course, gives rise to a basic
question about the meaning of the Lorentz symmetry. This paper gives enough
simple and transparent arguments, which should convince that usual Lorentz
covariant description refers, first of all, to the \textit{preferred frame},
namely the \textit{rest frame of the observer}. Such frame is the only one
where all physical measurements can be done. Furthermore, there is even no
need to touch on gravity issue since, as it will be shown, the relativity
aspect naturally emerges as a consequence of the fundamental postulates of
quantum mechanics, and thus it becomes a basic component of particle-wave
duality description. If one looks at special relativity as an integral part
of quantum mechanics, one finds that there is no need to interfere within
the well-known relativistic formulas to reconcile the absolute energy scale
with the Lorentz invariance, as well as, the relativity principle with the
existence of the \textit{preferred frame}. Indeed, one finds that special
relativity itself preserve the Planck length, however, the particles cannot
be seen any longer point-like but rather as extended quantum objects.

The structure of the paper is as follows. In Section 2 there are discussed
some general arguments testifying that relativistic description itself is
the \textit{preferred frame description}. In Section 3 it is shown how the
principles of quantum mechanics combined with the simplest (linear)
dispersion relation lead to the interval invariant relativity principle.
This analysis is preceded by a discussion which explains the two meanings of
time and an idea of \textit{relative scaling}. In Section 4 it is pointed
out the importance of a ``light-cone skeleton'' i.e. the two light cone
vectors, which transformed in a unitary way to the Mikowski frame, describe
four-momentum of particle with mass. Next, the diagonal form of the
Lorentz-boost transformation matrix is derived. The quantum-mechanical
foundations of space-time are discussed as well. In Section 5 a velocity
problem is considered within the context of Heisenberg uncertainty principle
and an arising picture of particle as an extended quantum object. This
provide us a new kinematical meaning for the Minkowski space-time. Finally,
in Section 6 a meaning of so-called proper time is reexamined.

\section{The relativity of inertial frames and the \textit{preferred frame}}

The special theory of relativity is based on two postulates: (1) the
postulate of relativity of motion and (2) the postulate of the constancy of
the speed of light. As already mentioned, there are the attempts to
challenge the latter. However, even if there are physical evidences of
varying the values of fundamental constants \cite{ec} the meaning of such
observations, although very instructive, is presumably not strong enough to
disrupt essentially traditional overtone of the special relativity. Though
the paradox of Magueijo and Smolin concerns mainly the momentum space
observation, it corresponds to the well known one of position space, namely,
the twin paradox. Therefore, the corrections introduced into the relativity,
based on the varying speed of light only, may not be up to the task of
refinement of complex relativity issue (especially in the case of ``pure
relativity analysis'', i.e. when additionally no gravity effects are taken
into the consideration).

Let us examine then the first relativity principle, which says \cite{Ein2}
``the laws by which the states of physical systems undergo change are not
affected, whether these changes of state be referred to the one or the other
of two systems of co-ordinates in uniform translatory motion''. This
principle firstly expresses our faith in universal character of nature laws.
However, our classical perception has led also to the classical conclusion
that ``there is a transformation group that converts measurements made by
one inertial observer to measurements made by another'' \cite{MS}. \ Note,
that this ``principle-conclusion'' cannot be directly verified by any
experimental technique, and has become the source of difficulties in
relativity interpretation. To approach the matter, let us consider first the
case of single observer and ask: is it possible to perform an experiment in
which the speed of light, or any other physical quantity, can be measured
outside the rest frame of the observer? According to Bell \cite{Bell2} ``the
only `observer' which is essential in orthodox practical quantum theory is
the inanimate apparatus which amplifies microscopic events to macroscopic
consequences''. But the `observer' always observes in `his' own rest frame
and there is no other possibility. So, the rest frame of the observer,
identified now with the laboratory frame, plays the role of the \textit{%
preferred frame,} where all the laws of nature are being discovered and
described. In the case of two observers the first relativity postulate could
be verified only when the two observers would measure the \textit{same
quantity }at the \textit{same time}. Whatever it means, it is clear that now
the observers have to be coupled. Thus, they are not independent any longer
and the case of two observers reduces simply to the single observer case.

Of course, the same physical phenomena may be observed by independent
observers placed in different inertial frames, and we know that their
theoretical predictions resulting from the same equations (although
established for different boundary conditions) agree. But this is just the
case when the\textit{\ relativity of inertial frames} manifests itself by
the fact that \textit{all (inertial) observers that make measurements in the
same conditions obtain the same experimental results}. In consequence, the
world seen by each of the observers looks the same. In other words, the
relativity principle reflects the most basic property of physical
observation, namely, its ability to reproduction. On the other hand, the
symmetry of derived wave equations (e.g. Maxwell equations) may include a
feature of \textit{preferred frame description} itself. Although, in
general, the choice of reference frame is free, so that there is no
preferred state of motion, form the experimental point of view the velocity
is not a purely relative quantity, despite what is commonly thought in a
spirit of Einstein interpretation.

\section{Two meanings of time and energy-momentum comparison scale}

The most natural definition of time characterizes time as a measure of pace
of observed changes. We will refer to the time used in such a meaning as to
the \textit{vital-time}. The time evolution of system should follow then a 
\textit{vital-time description.} The quantum mechanics, however, indicates
another time meaning. Due to the postulate of Planck, generalized later by
Einstein, the time may be also used as the energy measure by means of
inverse time units. Since the energy of particle characterized by the wave
of period $T$ (or frequency $\omega$) is 
\begin{equation}
\mathcal{E}=\frac{h}{T}=\hbar\omega,  \label{en}
\end{equation}
eq. (\ref{en}) may also serve as definition of the \textit{frozen-time }%
being the energy measure. Similarly, due to the postulate of de Broglie, the
value of particle momentum $\Pi$ corresponds to the wavelength $\lambda$
according to 
\begin{equation}
\Pi=\frac{h}{\lambda}.  \label{mom}
\end{equation}
These two fundamental postulates provide the rules for relating energy with
time and momentum with length. On the other hand, the momentum and energy,
similarly as the distance and time, are the two quantities that cannot be
directly compared, because they are expressed in different physical units.
Therefore, to compare them in the direct way one needs to introduce a
dimensional factor of velocity and thus set a \textit{comparison scale}. One
may ask then, whether the time that enters the description which expresses
time in length units is \textit{vital} or \textit{frozen?} An example
discussed below shows that these two time meanings, in general, are not
equivalent.

\subsection{The most simple dispersion relation}

One of the basic criterion for construction of particle wave equation is the
dispersion relation that must be obeyed. The simplest physical scenario
provides the situation in which the energy of particle (or quasiparticle) $%
\mathcal{E}$ is proportional to the momentum value $\Pi$. As mentioned, in
such a case the dispersion relation must involve some velocity $v$ and may
be put down in the form 
\begin{equation}
\mathcal{E}=\sigma v\Pi,  \label{en2}
\end{equation}
where $\sigma=\pm1,$ was introduced to allow for the positive and negative
energy values. Note, that arbitrarily chosen velocity value imposes the 
\textit{comparison scale} between the energy and momentum but, in general,
it cannot be identified with ``true'' particle velocity. On the other hand,
since the choice of $v$ is assumed to be free, one deals, in fact, with the
set of equivalent dispersion equations: $\mathcal{E}=\sigma v_{1}\Pi,$ $%
\mathcal{E}=\sigma v_{2}\Pi,...,$ differing just by the velocity factor.
This observation may be given in a simpler way by noticing that any velocity
value $v$ may be expressed by means of some preferred velocity $c$ and a
scaling factor $\eta>0$ as $v\equiv v_{\eta}=\eta^{2}c.$ It allows to
introduce the formula 
\begin{equation}
\frac{1}{\eta}\left( \frac{\mathcal{E}}{c}\right) =\sigma\eta\Pi,
\label{en3}
\end{equation}
in which given value $\eta$ corresponds to the one of dispersion equation
taken from the set.

Next, let us assume that it is possible to find a physical system
characterized by the linear energy-momentum dependence and the quasiparticle
excitations, which always propagate with a constat velocity $c$ in a given
medium (no matter how big such excitations are). In this case, the most
natural dispersion relation is given by eq. (\ref{en3}) for $\eta=1$. In
effect, such an idealized system may serve as the reference one to set the 
\textit{mappings} for energy 
\begin{equation}
\mathcal{E}\rightarrow\frac{\mathcal{E}}{c}=\Pi_{0},  \label{map1}
\end{equation}
and time 
\begin{equation}
t\rightarrow x_{0}=ct.  \label{map2}
\end{equation}
Obviously, such idealized system exists and is well-known, so it will be
simply called the \textit{photon system}. Note, that eqs. (\ref{en}), (\ref
{mom}) and (\ref{en2}) for $\sigma=1$ and $\eta=1$ yield the classical
dispersion relation $\lambda=cT,$ valid for single electromagnetic
monochromatic plane wave. Thus, naively speaking, in the time period $T$\ a
photon makes the distance 
\begin{equation}
\Delta x_{1}=\lambda=cT=\Delta x_{0}.  \label{lt}
\end{equation}
Relations (\ref{lt})\ show that in the case of the \textit{photon system}
the time interval $\Delta x_{0}/c$ has clear kinematical and dynamical
meaning, so that the \textit{vital} and \textit{frozen} time meanings can be
identified. However, when particle velocity depends on energy (or momentum)
the relationship between the \textit{vital} and \textit{frozen} time
intervals is unknown, and the equivalence between the two time meanings no
longer holds. Of course, one may argue that if particle velocity is known,
then for a given \textit{frozen }time interval the corresponding \textit{%
vital }one is known too. But the key point is that in the problems based on
momentum-space (position-space) description the velocity (energy) is a
quantity, which in principle is unknown. Roughly speaking, quantum effects
hidden under the cover of the Heisenberg uncertainty principle are the
reason for that. Note, that in order to establish in experimental way the
relationship between particle velocity and energy, one has to make the
simultaneous and independent measurement of both quantities. In particular,
the velocity has to be established in time-of-flight (TOF) method. Such
measurement, however, must violate somehow the Heisenberg uncertainty
principle 
\begin{equation}
\Delta E\Delta t\gtrsim h,  \label{un}
\end{equation}
where $\Delta t$ and $\Delta E$ are the uncertainties of time and energy
related to the measurement. We will return to this issue latter, when a
generalized form of the Heisenberg principle will be discussed. Note also
that Heisenberg uncertainty principle does not concern the \textit{photon
system} because of the postulate of the constancy of the speed of light, so
the \textit{photon system} seems to be favored once again.

Elementary analysis given above comes to the conclusion that the notion of
space-time refers to the space, which is reciprocal to energy-momentum one,
or in other words, that both spaces of momentum and position play the role
of \textit{reciprocal spaces }in traditional relativity formulation.
Therefore, the time that enters the space-time description, in general, may
be identified only with the \textit{frozen time}.

\subsection{The energy-momentum and time-distance \textit{relative scaling}}

We now consider the meaning of the scaling factor $\eta $ introduced in (\ref
{en3}). For $\eta =1$ one obtains the \textit{photon} dispersion relation 
\begin{equation}
\Pi _{0}=\sigma \Pi .  \label{mom2}
\end{equation}
In the case of $\eta \neq 1$, eq. (\ref{en3}) can be still written in the
same form 
\begin{equation}
\Pi _{0}^{\prime }=\sigma \Pi ^{\prime },  \label{mom22}
\end{equation}
where 
\begin{equation}
\Pi _{0}^{\prime }=\frac{1}{\eta }\Pi _{0}\text{ \ \ and \ \ }\Pi ^{\prime
}=\eta \Pi ,  \label{skal}
\end{equation}
are the values of energy and momentum considered, however, in the frame
which axes have been \textit{relatively re-scaled} with respect to the
reference frame given for $\eta =1.$ Similarly, according to (\ref{en}), (%
\ref{mom}) and (\ref{lt}) the scaling conditions (\ref{skal}) written in
space-time notation are: 
\begin{equation}
\Delta x_{0}^{\prime }=\eta \Delta x_{0}\text{ \ \ and \ \ }\Delta
x_{1}^{\prime }=\frac{1}{\eta }\Delta x_{1}.  \label{tr0}
\end{equation}
So, for $\eta >1,$ the transformations (\ref{tr0})\ may be called ``time
dilatation'' and ``length contraction''. We note also that dispersion eqs. (%
\ref{mom2}) and (\ref{mom22}) are in relation 
\begin{equation}
\Pi _{0}^{2}-\mathbf{\Pi }^{2}=\Pi _{0}^{\prime 2}-\mathbf{\Pi }^{\prime
2}=0,  \label{Linv1}
\end{equation}
which written in space-time notation takes the familiar form 
\begin{equation}
\left( \Delta x_{0}\right) ^{2}-\left( \Delta x_{1}\right) ^{2}=\left(
\Delta x_{0}^{\prime }\right) ^{2}-\left( \Delta x_{1}^{\prime }\right) ^{2}%
\text{ }=0.  \label{Linv2}
\end{equation}
It is obvious that the formalism based on arbitrarily chosen energy-momentum
(or time-distance) \textit{comparison scale} has to be covariant relative to
the change of this scale. The scaling factor $\eta $ that sets the \textit{%
comparison scale,} plays a role of a ``master-parameter'' in homogenous
Lorentz group. Namely, it will be shown that it ``splits'' into three boost
parameters.

\section{The \textit{Lorentz-boost} transformation in diagonal form}

The homogenous Lorentz group has six parameters, where three of them refer
to the subgroup of three-dimensional rotations. Currently we neglect the
very Euclidean aspect of the Lorentz transformations and concentrate only on
their most crucial \textit{boost-part}, which is going to be reduced to
diagonal form. We start from one-dimensional analysis by introducing a
concept of \textit{bimomentum,} helpful in description of particle mass. The
generalization into three dimensions will be straightforward. We start,
however, from pointing out some basic and important features hidden in
light-cone frame description.

\subsection{\textit{Bimomentum} and introduction of mass}

The \textit{photon system, }discussed already,\textit{\ }was given as an
example of an idealized system in description of which the \textit{vital}
and \textit{frozen} time meanings can be identified. In general, such an
equivalence of both time meanings applies to relativistic descriptions of
zero-mass fields, i.e.: the free electromagnetic field, free scalar field
and Weyl fields for spin-half particle. All these fields are characterized
by the same light-cone dispersion relation 
\begin{equation}
\Pi _{0}=\pm \left| \mathbf{\Pi }\right| ,  \label{phd}
\end{equation}
so that, their quasiparticle excitations are assumed to propagate always
with the same velocity $c$. Since the algebraic structure of these
(massless) fields now is of no importance, we will call them all the \textit{%
photon fields}. As already noticed, in the case of material particle, when
velocity is energy-dependent, the \textit{vital }meaning of time in
space-time description, in general, brakes down. Thus, the understanding of
correlation between \textit{vital} and \textit{frozen} time intervals seems
to be the key issue in encompassing of relativity matter.

The light-cone frame description, of course, goes much beyond the
formulation of dispersion relation for massless particles. It is well-known
the usefulness of light-cone frame description in quantization of strings in
the string theory \cite{string}. A characteristic feature of this light-cone
frame description is, that it appears in problems where particle mass comes
not as a feature of wave equation but rather its solutions. Indeed, another
example provides so-called covariant harmonic oscillator, discussed widely
by Kim and Noz beginning with \cite{kim73}. The issue of the covariant
harmonic oscillator originally was considered in a context of description of
relativistic hadrons, but, in general, one may say that it touches a problem
of description of composite quantum structures. Indeed, it was shown that
the wave equation given by Feynman, Kislinger and Ravndal \cite{F1},
proposed to describe a self-interacting hadron, separates into the
Klein-Gordon equation and the covariant harmonic oscillator one \cite{kim75}%
. As a result, the Klein-Gordon particles have masses which correspond
different solutions of the covariant harmonic oscillator equation.

In other words, the plane-wave properties of Klein-Gordon particles have
their origin in a quantum structure of bounded (particle) states. An arising
picture of particle-wave duality coming from above is rather clear.
Furthermore, although the dispersion relation obeyed by the Klein-Gordon
plane-waves is not the light-cone one, it will be shown that it has a
unitary light-cone ``equivalent''. On the other hand, also it will be shown,
that the ``size'' of this oscillatory-like particle may be placed at the
same light-cone ``structure''. In other words, one may indicate some \textit{%
light-cone skeleton} that unifies a description of particle-wave duality.

Since the main purpose of this paper is only an elementary analysis of the
relativity issue within the framework of the postulates of quantum
mechanics, some quantum-mechanical aspects related to the solutions of
covariant harmonic oscillator \cite{Kbook}, will serve us as an illustration
to the presented ideas, whereas the field theory analysis concerning this
issues will be given separately \cite{PK}.

To approach the matter, we first consider a concept of \textit{bimomentum }%
which makes possible to represent the massive and massless particle states
by means of the latter ones only.

The \textit{bimomentum} $\binom{\Pi_{0}}{\Pi_{1}}$ is defined as the
two-component light-cone momentum vector, which components may be referred
to two different solutions of \textit{photon} (i.e. massless) equation in
the following way: the first component $\Pi_{0}$ is the energy of the first 
\textit{photon}, whereas the second component $\Pi_{1}$ is the momentum of
the second \textit{photon}. The \textit{bivector} represents then a \textit{%
two-photon state}. Thus, the two \textit{light-cone photons} are assumed to
propagate along the same real space axis. For the purpose of this paper it
is enough to limit the analysis to the case where both \textit{photons} have
positive energies.

Let us next consider a unitary transformation of \textit{bimomentum} into 
\textit{effective bimomentum }defined as 
\begin{equation}
\left( 
\begin{array}{c}
p_{0} \\ 
p_{1}
\end{array}
\right) =\left( 
\begin{array}{cc}
cos\alpha & -sin\alpha \\ 
sin\alpha & cos\alpha
\end{array}
\right) \left( 
\begin{array}{c}
\Pi_{0} \\ 
\Pi_{1}
\end{array}
\right) ,  \label{biv01}
\end{equation}
where $p_{0}$ and $p_{1}$ are assumed to be the energy and momentum of
physically observed state, i.e. in a frame other then the light-cone one. 
\textit{\ }For particular choice of $\alpha=45^{0}$ we will call this frame
the Minkowski frame.\ Thus, we limit our discussion to the case 
\begin{equation}
\left( 
\begin{array}{c}
p_{0} \\ 
p_{1}
\end{array}
\right) =\left( 
\begin{array}{cc}
\frac{1}{\sqrt{2}} & \frac{-1}{\sqrt{2}} \\ 
\frac{1}{\sqrt{2}} & \frac{1}{\sqrt{2}}
\end{array}
\right) \left( 
\begin{array}{c}
\Pi_{0} \\ 
\Pi_{1}
\end{array}
\right) .  \label{biv}
\end{equation}
This transformation corresponds to the one introduced by Dirac \cite{dir2}
and has been already used to demonstrate the changes in four-momentum
distribution of boosted covariant harmonic oscillator ground state \cite
{kim77}.

In three dimensional real space the \textit{effective bimomentum} $\binom{%
p_{0}}{p_{1}}$ must correspond to the gauge Minkowski four-momentum, which
two components equal zero. As an example of transformation (\ref{biv}) it is
advisable to consider the two generic cases: (i) $\Pi_{1}=0,$ and (ii) $%
\Pi_{1}=-\Pi_{0}.$

In the first case the \textit{bimomentum} $\binom{\Pi_{0}}{\Pi}$ represents
a single (Minkowski) \textit{photon}, since due to (\ref{biv}) 
\begin{equation}
\left( 
\begin{array}{c}
\Pi_{0} \\ 
\Pi_{1}
\end{array}
\right) \underset{\Pi_{1}=0}{\longrightarrow}\left( 
\begin{array}{c}
\Pi_{0}/\sqrt{2} \\ 
\Pi_{0}/\sqrt{2}
\end{array}
\right) .  \label{biv0}
\end{equation}

In the second case, let us put 
\begin{equation*}
\Pi_{1}=\frac{mc}{\sqrt{2}}=-\Pi_{0},
\end{equation*}
which gives 
\begin{equation}
\text{\ }\left( 
\begin{array}{c}
\Pi_{0} \\ 
\Pi_{1}
\end{array}
\right) \underset{\Pi_{0}=-\Pi=\frac{mc}{\sqrt{2}}}{\longrightarrow}\left( 
\begin{array}{c}
mc \\ 
0
\end{array}
\right) .  \label{biv1}
\end{equation}
Thus, now the \textit{effective bimomentum }corresponds to the state of zero
momentum and non-vanishing energy $E_{0}=c\Pi_{0}\equiv mc^{2}$, i.e. the
ground state energy. One finds then, that the concept of \textit{bimomentum}
allows us to combine in a unitary way the two massless \textit{photon states}
into one \textit{effective state,} which may be either massless or massive.

\subsection{The \textit{relative scaling} and Lorentz symmetry}

It was argued that the scaling parameter $\eta$ (\ref{en3})\ was to be
considered as the parameter of \textit{preferred frame} \textit{description. 
}It was shown also that the change of $\eta,$ what corresponds to the change
of energy-momentum (\ref{skal})\ or time-distance (\ref{tr0}) \textit{%
comparison scale, }preserves the form of the \textit{photon} dispersion
relation (\ref{mom2}). We now consider the transformation of \textit{%
effective bimomentum} induced by the \textit{relative scaling }of \textit{%
bimomentum }components.

Let us assume that eq. (\ref{biv})\ is written in the frame given for $%
\eta=1.$ Thus, in the frame for which $\eta\neq1$ the form of \textit{%
effective bimomentum} is given by 
\begin{equation}
\left( 
\begin{array}{c}
p_{0}^{\prime} \\ 
p_{1}^{\prime}
\end{array}
\right) =\left( 
\begin{array}{cc}
\frac{1}{\sqrt{2}} & \frac{-1}{\sqrt{2}} \\ 
\frac{1}{\sqrt{2}} & \frac{1}{\sqrt{2}}
\end{array}
\right) \left( 
\begin{array}{c}
\Pi_{0}^{\prime} \\ 
\Pi_{1}^{\prime}
\end{array}
\right) ,  \label{bivp}
\end{equation}
where, due to (\ref{skal}), the coordinates of new (primed)\ and old
(unprimed) \textit{bimomenta }fulfil 
\begin{equation}
\left( 
\begin{array}{c}
\Pi_{0}^{\prime} \\ 
\Pi_{1}^{\prime}
\end{array}
\right) =\left( 
\begin{array}{cc}
\eta & 0 \\ 
0 & 1/\eta
\end{array}
\right) \left( 
\begin{array}{c}
\Pi_{0} \\ 
\Pi_{1}
\end{array}
\right) .  \label{skalm}
\end{equation}
By combining equations (\ref{biv}), (\ref{bivp}) and (\ref{skalm}) one finds
that relationship between the coordinates of \textit{\ bimomenta }in
different Minkowski frames is given by the Lorentz transformation formulas 
\begin{equation}
\left( 
\begin{array}{c}
p_{0}^{\prime} \\ 
p_{1}^{\prime}
\end{array}
\right) =\left( 
\begin{array}{cc}
cosh\xi & sinh\xi \\ 
sinh\xi & cosh\xi
\end{array}
\right) \left( 
\begin{array}{c}
p_{0} \\ 
p_{1}
\end{array}
\right) \text{ \ where }\xi=ln\eta,  \label{Lor1}
\end{equation}
This points out to the dependence on the Lorentz symmetry with the freedom
of scaling $\eta$.

More precisely, the transformation (\ref{Lor1}) may be understood as: (1)
the \textit{passive} one, when both \textit{bi-momenta }$\binom{p_{0}}{p_{1}}
$ and $\binom{p_{0}^{\prime}}{p_{1}^{\prime}}$ refer to the one physical
state but are described in two different frames (i.e. in the frames based on
different \textit{comparison scales}), or (2) as the \textit{active} one
when both \textit{bi-momenta} are considered in the same \textit{preferred
frame} but refer to two different \textit{physical states}. We will say that
in the case (1) $\eta$ is the \textit{passive} parameter, whereas in the
case (2) $\eta$ is the \textit{active }one\textit{.} So, in the latter case
the $\eta-parametrization$ does not concern the frame characteristics but a
dynamical feature of the state. We write this down in the explicit form 
\begin{equation}
\left( 
\begin{array}{c}
p_{0} \\ 
p_{1}
\end{array}
\right) =\left( 
\begin{array}{cc}
\gamma & \gamma\beta \\ 
\gamma\beta & \gamma
\end{array}
\right) \left( 
\begin{array}{c}
mc \\ 
0
\end{array}
\right) ,  \label{Lor2}
\end{equation}
where $\gamma=cosh\xi$ and $\gamma\cdot\beta=sinh\xi$. Formally, the
transformation matrixes given in (\ref{Lor1})\ and (\ref{Lor2})\ are
identical. However, the parameters $\gamma$ and $\beta$ (which kinematical
meaning will be discussed latter) were introduced to emphasis that their
values need to be considered with reference to the ground state energy.

Since the ground state energy is characterized by zero momentum, the
transformation (\ref{Lor2}),\ generalized into the three-dimensional case,
should take the form 
\begin{equation}
\left( 
\begin{array}{c}
p_{0} \\ 
p_{1} \\ 
p_{2} \\ 
p_{3}
\end{array}
\right) =\left( 
\begin{array}{cccc}
\gamma & \gamma\beta & 0 & 0 \\ 
\gamma\beta & \gamma & 0 & 0 \\ 
0 & 0 & 1 & 0 \\ 
0 & 0 & 0 & 1
\end{array}
\right) \left( 
\begin{array}{c}
mc \\ 
0 \\ 
0 \\ 
0
\end{array}
\right) ,  \label{Lor3}
\end{equation}
which describes a boost in the $x$ direction ($p_{2}=p_{3}=0$). The matrix
form of eq. (\ref{Lor3}) is 
\begin{equation}
P_{x}=A_{0}\text{ }mc.  \label{Lor3s}
\end{equation}
To describe a boost in any direction one needs to rotate the initial frame,
first (let say) in $xy$ and next in $yz$ plane. In the new frame eq. (\ref
{Lor3s}) takes the form 
\begin{equation}
R_{yz}R_{xy}P_{x}=R_{yz}R_{xy}A_{0}R_{xy}^{-1}R_{yz}^{-1}R_{yz}R_{xy}mc,
\label{Lor3r}
\end{equation}
where 
\begin{equation}
R_{xy}=\left( 
\begin{array}{cccc}
1 & 0 & 0 & 0 \\ 
0 & cos\varphi & -sin\varphi & 0 \\ 
0 & sin\varphi & cos\varphi & 0 \\ 
0 & 0 & 0 & 1
\end{array}
\right) ,\text{ }R_{yz}=\left( 
\begin{array}{cccc}
1 & 0 & 0 & 0 \\ 
0 & 1 & 0 & 0 \\ 
0 & 0 & cos\psi & -sin\psi \\ 
0 & 0 & sin\psi & cos\psi
\end{array}
\right) ,  \label{rotM}
\end{equation}
so that, the angles $\varphi$ and $\psi$ describe the rotations around the $%
z $ and $x$ axes respectively. The explicit form of eq. (\ref{Lor3r}) is
then given by 
\begin{equation}
\left( 
\begin{array}{c}
E/c \\ 
p_{x} \\ 
p_{y} \\ 
p_{z}
\end{array}
\right) =\left( 
\begin{array}{cccc}
\gamma & \gamma\beta_{1} & \gamma\beta_{2} & \gamma\beta_{3} \\ 
\gamma\beta_{1} & 1+\frac{(\gamma-1)}{\beta^{2}}\beta_{1}^{2} & \frac
{(\gamma-1)}{\beta^{2}}\beta_{1}\beta_{2} & \frac{(\gamma-1)}{\beta^{2}}%
\beta_{1}\beta_{3} \\ 
\gamma\beta_{2} & \frac{(\gamma-1)}{\beta^{2}}\beta_{1}\beta_{2} & 1+\frac{%
(\gamma-1)}{\beta^{2}}\beta_{2}^{2} & \frac{(\gamma-1)}{\beta^{2}}%
\beta_{2}\beta_{3} \\ 
\gamma\beta_{3} & \frac{(\gamma-1)}{\beta^{2}}\beta_{1}\beta_{3} & \frac{%
(\gamma-1)}{\beta^{2}}\beta_{2}\beta_{3} & 1+\frac{(\gamma-1)}{\beta ^{2}}%
\beta_{3}^{2}
\end{array}
\right) \left( 
\begin{array}{c}
mc \\ 
0 \\ 
0 \\ 
0
\end{array}
\right) ,  \label{Lor4}
\end{equation}
where $\beta_{1}/\beta=cos\varphi$, $\beta_{2}/\beta=sin\varphi\cdot cos\psi$
and $\beta_{3}/\beta=sin\varphi\cdot sin\psi.$ This leads to the formulas
for energy $E=\gamma mc^{2}$ and momentum $\mathbf{p}=\gamma\mathbf{\beta}$,
where $\mathbf{\beta=}(\beta_{1},\beta_{2},\beta_{3}).$ Eq. (\ref{Lor4}) may
be written as 
\begin{equation}
P=\mathbf{A}\text{ }m.  \label{Lor4s}
\end{equation}
To recollect, the matrix $\mathbf{A}$ represents the \textit{Lorentz-boost}
transformation. Traditionally it is derived by means of the boost generators
of the Lorentz group and the Taylor expansion \cite{Jack}. The way it was
constructed now allows us to express the matrix $\mathbf{A}$ in the form 
\begin{equation}
\mathbf{A}=\mathbf{RU\Lambda}_{\eta}\mathbf{U}^{-1}\mathbf{R}^{-1},
\label{Lor4A}
\end{equation}
where 
\begin{equation}
\mathbf{\Lambda}_{\eta}=\left( 
\begin{array}{cccc}
\eta & 0 & 0 & 0 \\ 
0 & 1/\eta & 0 & 0 \\ 
0 & 0 & 1 & 0 \\ 
0 & 0 & 0 & 1
\end{array}
\right) ,\text{ }\mathbf{U}=\left( 
\begin{array}{cccc}
\frac{1}{\sqrt{2}} & -\frac{1}{\sqrt{2}} & 0 & 0 \\ 
\frac{1}{\sqrt{2}} & \frac{1}{\sqrt{2}} & 0 & 0 \\ 
0 & 0 & 1 & 0 \\ 
0 & 0 & 0 & 1
\end{array}
\right) ,\text{ }\mathbf{R}=R_{yz}\circ R_{xy}.  \label{rotA}
\end{equation}
Since neither of the matrixes $\mathbf{R}$ nor $\mathbf{U}$ is singular the
diagonal form of the matrix $\mathbf{A}$ is given by $\mathbf{\Lambda}%
_{\eta} $. So, it was shown that, although there are three different boosts
parameters, their origin is provided by the one scaling factor $\eta,$
established on a ground of purely quantum-mechanical considerations.

\subsection{Quantum-mechanical foundations of the Minkowski space}

The notion of space-time origins from the classical analysis of the
electromagnetic field. Currently we show that the space-time naturally
emerges as the \textit{reciprocal} to energy-momentum one.

To recognize the ``reciprocal dependence'' of position and momentum spaces
we start again from the momentum space by considering the scaling
transformation (\ref{skalm}) acting on \textit{bimomentum} of the ground
state, namely 
\begin{equation}
\left( 
\begin{array}{c}
\Pi _{0} \\ 
\Pi _{1}
\end{array}
\right) =\left( 
\begin{array}{cc}
\eta & 0 \\ 
0 & 1/\eta
\end{array}
\right) \left( 
\begin{array}{c}
\frac{1}{\sqrt{2}}mc \\ 
-\frac{1}{\sqrt{2}}mc
\end{array}
\right) ,  \label{skelpG}
\end{equation}
where $\eta $ plays the role of \textit{active} parameter. Thus, according
to (\ref{biv}), the components of \textit{bi-momentum} related to \textit{%
bimomentum} $\binom{\Pi _{0}}{\Pi _{1}}$ in (\ref{skelpG}) may be expressed
via \textit{active} $\eta $ as 
\begin{equation}
p_{0}=\frac{1}{2}\left( \eta +\frac{1}{\eta }\right) mc,\text{ \ \ }p_{1}=%
\frac{1}{2}\left( \eta -\frac{1}{\eta }\right) mc.\text{\ }  \label{activm}
\end{equation}
Note, that the coordinates of both \textit{bimomenta }(\ref{skelpG}) (and
thus the ground state energy $mc^{2}$) refer to the \textit{preferred frame
description,} for which the \textit{passive} $\eta $ is assumed to be equal
one. One easily finds that $\Pi _{0}$ and $\Pi _{1}$ given in (\ref{skelpG})
satisfy the condition 
\begin{equation}
\left( \Pi _{0}-\Pi _{1}\right) ^{2}-\left( \Pi _{0}+\Pi _{1}\right)
^{2}=\left( mc\right) ^{2},  \label{Linv3}
\end{equation}
which is invariant with respect to the choice of $\eta .$ As already
mentioned, the mappings (\ref{map1}) and (\ref{map2})\ allow us to write
down the components of \textit{bimomenta }in terms of the wavelengths, i.e. 
\textit{generalized Compton wavelength} $\lambda _{1}$ and de Broglie
wavelength $\lambda _{2}$ defined as 
\begin{equation}
\Pi _{0}=\frac{h}{\lambda _{1}}\text{ \ \ \ and \ \ }\Pi _{2}=-\frac{h}{%
\lambda _{2}}.  \label{pass}
\end{equation}
Relations (\ref{pass}) provide then a basis for a transition from the
momentum to position space. Thus, in position space one finds 
\begin{equation}
\left( 
\begin{array}{c}
\lambda _{1} \\ 
\lambda _{2}
\end{array}
\right) =\left( 
\begin{array}{cc}
1/\eta & 0 \\ 
0 & \eta
\end{array}
\right) \left( 
\begin{array}{c}
\sqrt{2}\lambda _{0} \\ 
\sqrt{2}\lambda _{0}
\end{array}
\right) ,  \label{Gskelp}
\end{equation}
where due to (\ref{skelpG}) 
\begin{equation}
\lambda _{0}=\frac{h}{mc},  \label{cpt}
\end{equation}
is the ``proper'' Compton wavelength.\ Eq. (\ref{Linv3})\ written in its
reciprocal version, takes the form 
\begin{equation}
\left( \frac{\lambda _{1}+\lambda _{2}}{2}\right) ^{2}-\left( \frac{\lambda
_{1}-\lambda _{2}}{2}\right) ^{2}=2\left( \frac{h}{mc}\right) ^{2}=const,
\label{Linv4}
\end{equation}
which shows that the right hand side of eq. (\ref{Linv4}) is (again) an $%
\eta -$scaling invariant.

Eqs. (\ref{skelpG})\ and (\ref{Gskelp}) are the light-cone equations.
Similarly, like in the case of momentum space analysis, one may introduce
the Minkowski frame representation for the wavelengths $\lambda_{1}$ and $%
\lambda_{2},$ namely 
\begin{equation}
\left( 
\begin{array}{c}
x_{0} \\ 
x_{1}
\end{array}
\right) =\left( 
\begin{array}{cc}
\frac{1}{\sqrt{2}} & \frac{1}{\sqrt{2}} \\ 
\frac{-1}{\sqrt{2}} & \frac{1}{\sqrt{2}}
\end{array}
\right) \left( 
\begin{array}{c}
\lambda_{1} \\ 
\lambda_{2}
\end{array}
\right) .  \label{pos1}
\end{equation}
Thus, for any two pairs of variables $\left\{ x_{0},x_{1}\right\} $ and $%
\left\{ x_{0}^{\prime},x_{1}^{\prime}\right\} $ (corresponding to the two
different values of $\eta$ and $\eta^{\prime}$) it must occur 
\begin{equation}
x_{0}^{2}-x_{1}^{2}=x_{0}^{\prime2}-x_{1}^{\prime2}=2\left( \frac{h}{mc}%
\right) ^{2}=const.  \label{delta}
\end{equation}
The connection between the quantum mechanics and special relativity is
obvious. Eqs. (\ref{delta})\ states the equivalent of eqs. (\ref{Linv2}) in
the case of non-zero mass. The different definitions of time and space
intervals (\ref{lt})\ and (\ref{pos1}) (distinguishing the massless and
massive cases) cause that expression (\ref{delta}) does not reduce to (\ref
{Linv2}) in the limit $m\rightarrow0$. Note, that the quantum aspect of the
approach appears explicitly just along with the introduction of mass.

The generalization for three dimensions is similar to the corresponding
procedure in the momentum space, but now we are mainly interested in
transformation (\ref{Lor1}) as that \textit{passive} one. Let us consider a
particle state of energy $p_{0}$ and momentum $\mathbf{p}=(p_{1},0,0).$ The
quantities $p_{0}$ and $p_{1}$ can be expressed by means of \textit{active} $%
\eta$ (\ref{activm}). On the other hand $p_{0}$ and $p_{1},$ through (\ref
{biv}) and (\ref{pass}) can also be expressed in terms of reciprocal
quantities $x_{0}$ and $x_{1}$ given in (\ref{pos1}). One may ask then about
the space-time transformation, which corresponds to the \textit{active}
transition $\binom{p_{0}}{p_{1}}$ $\underset{\eta_{1}\rightarrow\eta_{2}}{%
\longrightarrow}$ $\binom{p_{0}^{\prime}}{p_{1}^{\prime}}$. One easily finds
that this \textit{passive} transformation $\{x_{0,}x_{1,}x_{2,}x_{3}\}%
\rightarrow\{x_{0,}^{\prime}x_{1,}^{\prime}x_{2,}^{\prime}x_{3}^{\prime}\}$
takes the form 
\begin{equation}
\left( 
\begin{array}{c}
x_{0}^{\prime} \\ 
x_{1}^{\prime} \\ 
x_{2}^{\prime} \\ 
x_{3}^{\prime}
\end{array}
\right) =\left( 
\begin{array}{cccc}
cosh\xi & -sinh\xi & 0 & 0 \\ 
-sinh\xi & cosh\xi & 0 & 0 \\ 
0 & 0 & 1 & 0 \\ 
0 & 0 & 0 & 1
\end{array}
\right) \left( 
\begin{array}{c}
x_{0} \\ 
x_{1} \\ 
x_{2} \\ 
x_{3}
\end{array}
\right) \text{ ,}  \label{Lor5}
\end{equation}
where\ $\xi=ln(\eta_{1}/\eta_{2})$. \ In more general case, when the
real-space axes of both frames are parallel but direction of the boost
direction $\widehat{\mathbf{\beta}}$ does not match any of the axes
directions, the transformed form of eq. (\ref{Lor5}) is given by 
\begin{equation}
\left( 
\begin{array}{c}
x_{0}^{\prime} \\ 
x_{1}^{\prime} \\ 
x_{2}^{\prime} \\ 
x_{3}^{\prime}
\end{array}
\right) =\left( 
\begin{array}{cccc}
\gamma & -\gamma\beta_{1} & -\gamma\beta_{2} & -\gamma\beta_{3} \\ 
-\gamma\beta_{1} & 1+\frac{(\gamma-1)}{\beta^{2}}\beta_{1}^{2} & \frac
{(\gamma-1)}{\beta^{2}}\beta_{1}\beta_{2} & \frac{(\gamma-1)}{\beta^{2}}%
\beta_{1}\beta_{3} \\ 
-\gamma\beta_{2} & \frac{(\gamma-1)}{\beta^{2}}\beta_{1}\beta_{2} & 1+\frac{%
(\gamma-1)}{\beta^{2}}\beta_{2}^{2} & \frac{(\gamma-1)}{\beta^{2}}%
\beta_{2}\beta_{3} \\ 
-\gamma\beta_{3} & \frac{(\gamma-1)}{\beta^{2}}\beta_{1}\beta_{3} & \frac{%
(\gamma-1)}{\beta^{2}}\beta_{2}\beta_{3} & 1+\frac{(\gamma-1)}{\beta ^{2}}%
\beta_{3}^{2}
\end{array}
\right) \left( 
\begin{array}{c}
X_{0} \\ 
X_{1} \\ 
X_{2} \\ 
X_{3}
\end{array}
\right) ,  \label{Lor6}
\end{equation}
where 
\begin{equation}
\left( 
\begin{array}{c}
X_{0} \\ 
X_{1} \\ 
X_{2} \\ 
X_{3}
\end{array}
\right) =R_{yz}R_{xy}\left( 
\begin{array}{c}
x_{0} \\ 
x_{1} \\ 
x_{2} \\ 
x_{3}
\end{array}
\right) .
\end{equation}
\ This ensures, of course, that the condition of interval invariance 
\begin{equation}
\left( x_{0}\right) ^{2}-\left( x_{1}\right) ^{2}-\left( x_{2}\right)
^{2}-\left( x_{3}\right) ^{2}=\left( x_{0}^{\prime}\right) ^{2}-\left(
x_{1}^{\prime}\right) ^{2}-\left( x_{2}^{\prime}\right) ^{2}-\left(
x_{3}^{\prime}\right) ^{2}=const,  \label{Linv5}
\end{equation}
is fulfilled. Note, that this condition being the principle of relativity
formulation, due to (\ref{Linv2})\ and (\ref{delta}), now, finds a
quantum-mechanical basis.

In this section it was shown how the Minkowski position-space appears as
reciprocal to the energy-momentum one. In the following section we consider
a kinematical meaning of description in Minkowski space-time.

\section{A velocity problem and Heisenberg uncertainty principle}

The textbook form of the Lorentz (space-time) transformations involves
velocity as a purely relative quantity. In currently discussed approach, the
notion of velocity refers only to the rest frame of the observer, so that
the velocity of observed quantum objects is no longer a relative quantity,
what changes essentially the meaning of the well-known transformation
formulas.

Before we discuss this issue, we start by considering another velocity
aspect, namely, its relationship with particle energy and momentum. As
already noticed, one needs to be very careful when uses the velocity notion
in the context of momentum space description. An example discussed below
reveals the related difficulty.

Derived in the previous section the transformation matrix $A_{0}$ (\ref
{Lor3s}) was given in terms of parameters 
\begin{equation}
\gamma=cosh\xi\text{ \ \ \ and \ \ }\gamma\cdot\beta=sinh\xi,  \label{ptex}
\end{equation}
where, due to (\ref{Lor1})\ and (\ref{Lor2}), the parameter $\xi$ is
directly related to the \textit{active} $\eta.$ The textbook form of
parameters (\ref{ptex}) provide the formulas 
\begin{equation}
\gamma=\frac{1}{\sqrt{1-w^{2}/c^{2}}}\text{ and \ }\beta=\frac{w}{c},
\label{wvelo}
\end{equation}
which involve some velocity $w$ interpreted as velocity of a particle. Eq. (%
\ref{Lor2}) may be then used to express the energy and momentum by means of
the velocity $w$ according to 
\begin{equation}
E=\frac{mc^{2}}{\sqrt{1-w^{2}/c^{2}}},\text{ \ \ \ }p=\frac{mw}{\sqrt
{1-w^{2}/c^{2}}}.  \label{Epfor}
\end{equation}
However, the parametrization of $\gamma$ and $\beta$ in terms of velocity $w$
is not unique. An alternative expressions can be found by introducing a new
velocity $v$ related to $w$ by the formula 
\begin{equation}
v=\frac{w}{\sqrt{1-w^{2}/c^{2}}},  \label{vw}
\end{equation}
which yields 
\begin{equation}
\gamma=\sqrt{1+v^{2}/c^{2}}\text{ \ and \ }\gamma\cdot\beta=\frac{v}{c}.
\label{vw2}
\end{equation}
As a result, the energy and momentum formulas (\ref{Epfor}) may be replaced
by the new ones 
\begin{equation}
E=mc^{2}\sqrt{1+v^{2}/c^{2}},\text{ \ \ }p=mv.  \label{Epfor2}
\end{equation}
One sees that the momentum (\ref{Epfor2}) has the classical form, whereas
the energy is the regular function of $v$ in the whole range. However, the
physical interpretation of expressions (\ref{Epfor2}) is rather troublesome,
since the velocities of magnitude grater then $c$ are, in general, not
observed. Thus, the arguments supporting the physical meaning of velocity $%
w, $ but not $v,$ are rather clear.

It is worthwhile to recollect now the two early experimental results that
are thought to confirm the interpretative foundations of special relativity,
namely the observations of high energy muons \cite{muons}, and the
measurement electron speed in Bertozzi experiment \cite{Berto}.

In the first case it was observed, that as a result of collisions of high
energy cosmic radiation protons with nucleus of the upper part atmosphere,
the pions are produced. These pions, because of its decay, become a source
of abundantly produced muons which state a major part of the secondary
cosmic radiation at the sea level. Thus, the distance made by such muon was
roughly $20$ $km.$ On the other hand, the mean time of muon life established
in its rest frame is $\tau\approx2\cdot10^{-6}$ $s$ . For a typical muon
energy of $3$ $GeV$, one may put $w\approx c,$ which yields the range $%
c\tau\approx600$ $m$ only. Thus, to explain this observation on the
assumption that velocity of real particle cannot exceed that of light, one
needs to assume that the life time of moving muon is increased relative to
its life time at rest. The complete explanation of this effect is ascribed
to pure geometrical property of the Minkowski space, called the time
dilatation effect. According to this, the passage of time in a moving frame
is less then in a stationary one.

In the case of Bertozzi experiment, the purpose was to verify the
correctness of formula (\ref{Epfor}). In this experiment the energy of
electron ($\lesssim$ 5 MeV) was measured in the calorimetric way and the
velocity by use TOF method. Indeed the correctness of formula (\ref{Epfor})
was confirmed. Nevertheless, both quantities i.e. energy and velocity were
established simultaneously and thus, as already noticed, the interpretation
of the experimental results obtained this way should have been somehow
referred to the Heisenberg uncertainty principle.

Discussed below quantum-mechanical approach to these problems make us
possible to avoid a confused (\textit{vital}) time relativity aspect, as
well as, to incorporate into the considerations the Heisenberg principle,
however, in its generalized form.

Let us assume that a particle seen by the apparatus looks more like an 
\textit{extended quantum object} than a material point. This might cause
that the effective \ (i.e. measured) time-of-flight interval $\Delta t_{w}$
is increased relative to the real one $\Delta t_{v}$ because of the
interaction between the electron and apparatus. The two (formally
equivalent) possible ways of energy and momentum parametrization (\ref{Epfor}%
)\ and (\ref{Epfor2}), allow us to postulate the mutual dependence between
both time intervals in the form 
\begin{equation}
\Delta l=w\Delta t_{w}=v\Delta t_{v},  \label{dist}
\end{equation}
where $\Delta l$ is a distance of particle flight, which may be identified
either with the total flight distance or just a part of it only (this
ambiguity will be explained below). Thus, combining (\ref{vw}) and (\ref
{dist}) one finds that the effective and real time intervals of particle
flight satisfy 
\begin{equation}
\Delta t_{w}=\frac{\Delta t_{v}}{\sqrt{1-w^{2}/c^{2}}}.  \label{tt}
\end{equation}
To see that suggested interpretation of (\ref{tt}) is reasonable, firstly,
let us assume that the quantum objects we are interested in are really
extended, i.e. have some space-time structures. Let us assume next, that we
know how such structure looks like, at least for a one particular particle
state. We will say that a particle has a \textit{model-shape} if extensions
of its ground state (in a way explained below) are characterized by the two
position light-cone vectors of the same magnitude 
\begin{equation}
\Lambda _{0}=\frac{1}{2}\lambda _{0},
\end{equation}
where $\lambda _{0}$ was given in (\ref{cpt}). Thus, due to (\ref{pos1}) if
the object moves, its ``light-cone shape'' changes from 
\begin{equation}
\left( 
\begin{array}{c}
\Lambda _{0} \\ 
\Lambda _{0}
\end{array}
\right) \overset{\text{to}}{\rightarrow }\left( 
\begin{array}{c}
\Lambda _{1} \\ 
\Lambda _{2}
\end{array}
\right) =\left( 
\begin{array}{c}
\frac{1}{\eta }\Lambda _{0} \\ 
\eta \Lambda _{0}
\end{array}
\right) .  \label{lc0}
\end{equation}
However, similarly like in the case of momentum space, we assume now that
physically observed quantities are not the light-cone ones but the
corresponding them Minkowski ``equivalents'', which form is given by 
\begin{equation}
\left( 
\begin{array}{c}
\Delta x_{0} \\ 
\Delta x_{1}
\end{array}
\right) =\left( 
\begin{array}{cc}
\frac{1}{\sqrt{2}} & \frac{1}{\sqrt{2}} \\ 
-\frac{1}{\sqrt{2}} & \frac{1}{\sqrt{2}}
\end{array}
\right) \left( 
\begin{array}{c}
\Lambda _{1} \\ 
\Lambda _{2}
\end{array}
\right) .  \label{lc1}
\end{equation}

The physical meaning of intervals $\Delta x_{0}$ and $\Delta x_{1}$seems to
be complement one another. Indeed, let us express these intervals by making
use of the formulas (\ref{lc1}), (\ref{lc0}) and (\ref{wvelo}). One easily
finds that 
\begin{align}
\Delta x_{0}& =\frac{1}{\sqrt{1-w^{2}/c^{2}}}\text{ }\lambda _{0},
\label{uc1} \\
\Delta x_{1}& =\frac{w/c}{\sqrt{1-w^{2}/c^{2}}}\text{ }\lambda _{0}.
\label{uc2}
\end{align}
Note, that $\Delta x_{0}\geqslant \Delta x_{1}$ and in particular, for $w$ $%
\lessapprox $ $c$, both interval are almost equal, but in the case of $w=0$, 
$\Delta x_{0}=\lambda _{0}$ whereas $\Delta x_{1}=0.$ This suggests that $%
\Delta x_{1}$ may be interpreted as the uncertainty of the particle center
(of mass) position inside some ``quantum region'' (which extension will be
given latter), and in a time interval resulting from (\ref{uc1}). Thus, one
may say that the uncertainty $\Delta x_{1}$ concerns a \textit{%
point-like-particle }position. If this \textit{point-like-particle} moves
inside the mentioned quantum region, it is useful to call such motion a
movement in a \textit{classical channel.} On the other hand, since $\Delta
x_{0}$ must correspond to particle time-like separation, let us assume that
it simply corresponds to the time of life of particle quantum state (do not
mistake with particle life time), for which the width (or uncertainty) of 
\textit{classical channel }is $\Delta x_{1}.$ Thus, it is useful to call $%
\Delta x_{0}$ the uncertainty of a \textit{quantum channel.}

To find the explicit dependence between the assumed life times of quantum
states for moving particle and particle at rest, let as put 
\begin{equation}
\lambda_{0}=c\Delta\tau\text{ \ and \ }\Delta x_{0}=c\Delta t,  \label{uc10}
\end{equation}
so that, the interval $\Delta\tau$ represents the time of life of the ground
state of the particle of \textit{model-shape.} Thus, by combining the
formulas (\ref{uc10}) and (\ref{uc1}) one obtains 
\begin{equation}
\Delta t=\frac{\Delta\tau}{\sqrt{1-w^{2}/c^{2}}},  \label{uc11}
\end{equation}
which corresponds to the formula (\ref{tt}). Currently, however, the
interval $\Delta t$ is to be interpreted as the life time of the quantum
state in which the uncertainty of \textit{point-like particle} position is 
\begin{equation}
\Delta x_{1}=v\text{ }\Delta\tau=w\Delta t,  \label{uc12}
\end{equation}
which, in turn reproduces the formula (\ref{dist}). The special case is, of
course, when the times of particle life and its given quantum state are the
same. In this case, given approach easily explains mentioned above time
dilatation effect.

Discussed approach also provides a new look at particle kinematics. Note,
that due to (\ref{uc12}) both time intervals $\Delta \tau $ and $\Delta t$
may serve as \textit{vital time} parameters. Seemingly, i.e. from the
classical point of view, only the interval $\Delta \tau $ would be
considered as the \textit{vital} one, since it directly concerns the assumed
continuous motion of \textit{point-like particle} \textit{\ }inside \textit{%
classical channel}, whereas the interval $\Delta t$ includes the time at
which, naively speaking, \textit{point-like particle} stay at rest, or
perhaps it is rather better to say, it oscillates around some stationary
point. Nevertheless, although the period $\Delta t$ encompasses the time at
which there is no propagation along the \textit{classical channel, }it may
play a role of \textit{vital time} parameter in a global scale, i.e. if one
considers a movement of the whole \textit{extended quantum object }at
effective velocity $w$ in a laboratory frame, or in other words in a \textit{%
quantum channel. }Therefore, neither \textit{classical} nor \textit{quantum }%
way of movement is smooth but must take a form of ``oscillatory'' or
``jump-like'' motion.

This, in turn, leads to a very important conclusion that considered \textit{%
extended quantum states} may be seen as temporarily localized in a given
observer rest frame. This just explains the assumption made above that $%
\Delta t_{w}=\Delta t.$ On the other hand, one finds that in Minkowski frame
identified with given laboratory frame, both time intervals $\Delta t$ and $%
\Delta\tau$ find their clear \textit{vital} meanings.

Another aspect of peculiar kind motion outlined above are the momentum
fluctuations involved in a mechanism of \textit{classical }and \textit{%
quantum channel }transmissions. This can be explained within the framework
of $\ $``updated'' Heisenberg uncertainty principle. Indeed, let us consider
again the intervals $\Delta x_{0}$ and $\Delta x_{1}$ (\ref{uc1}),(\ref{uc2}%
), but now let us write them down in a new but quite equivalent form 
\begin{align}
\Delta x_{0}& =\frac{h}{\Delta p}+\alpha ^{\prime }\frac{\Delta p}{h},
\label{uc3} \\
\Delta x_{1}& =\frac{h}{\Delta q}-\alpha ^{\prime }\frac{\Delta q}{h},
\label{uc4}
\end{align}
where $\alpha ^{\prime }=(h/mc)^{2}$ and 
\begin{equation}
\Delta p=\frac{h}{\Lambda _{2}},\text{ \ \ }\Delta q=\frac{h}{\Lambda _{1}}.
\label{uc5}
\end{equation}
The quantities $\Delta p$ and $\Delta q$ are to be interpret as the widths
of the momentum uncertainties for the \textit{extended }and \textit{%
point-like }quantum \textit{objects }just introduced above. In fact, the
light-cone relations (\ref{uc5}) are already known, since they have been
widely discussed in the context of the covariant harmonic oscillator
equation \cite{kim77},\cite{kimUC}. Therefore, there is no problem to
establish a correspondence between the introduced light-cone parameters $%
\Lambda _{1}$ and $\Lambda _{2}$ and the area of maximum density probability
distribution for the ground state of (boosted or un-boosted) covariant
harmonic oscillator. Furthermore, one easily recognize that $\Lambda _{1}$
and $\Lambda _{2}$ must correspond the size of the \textit{point-like} and 
\textit{extended quantum object }respectively. Actually, one may go even
further and show that the behavior of the ``shape'' of the whole \textit{%
extended} \textit{quantum object} (i.e. with the \textit{point-like object}
inside the extended one) is to be describe by O(3)-like group contraction
simultaneously to the Euclidean E(2) and cylindrical group \cite{kim90}.
This issue, however needs a separate treatment. \textit{\ }

Let us return to the eqs. (\ref{uc3}) and (\ref{uc4}). Since $\alpha
_{m}^{\prime}\geqslant\alpha^{\prime}=\left( l_{P}\right) ^{2}$ and $\Delta
x_{1}\leqslant\Delta x_{0},$ one easily finds that 
\begin{equation}
\Delta x_{0}\geqslant\frac{h}{\Delta p}+\alpha^{\prime}\frac{\Delta p}{h},
\label{uc6}
\end{equation}
and 
\begin{equation}
\Delta x_{1}\geqslant\frac{h}{\Delta q}-\alpha^{\prime}\frac{\Delta q}{h}.
\label{uc7}
\end{equation}
Relation (\ref{uc6}) is known as generalized Heisenberg uncertainty
principle and was proposed by Witten \cite{witten} in the context of duality
in string theory. The first observation that comes to mind now, is that this
duality should correspond a particle-wave duality. The second observation is
that the covariant harmonic oscillator may serve as a bridge between the
quantum mechanics, spacial relativity and the string theory. Indeed it is
hard to resist an impression that the open and closed strings somehow must
correspond to unbounded and bounded solutions appearing in the problem of
relativistic oscillator. But this issue, of course, goes much beyond the
scope of this paper. And finally, what is really now certain, is that
special relativity itself preserves the Planck length $l_{P}$, however, a
particle cannot be seen any longer a material point , but rather as an
extended quantum objects endowed with an internal structure.

At the end of this section, let us return to the problem of velocity meaning
in the Lorentz transformation formulas. Indeed, the transformation (\ref
{Lor5}) parametrized by means of the formulas (\ref{wvelo}) yields the known
expressions 
\begin{equation}
\text{\ }t^{\prime }=\frac{t-(w/c^{2})x}{\sqrt{1-w^{2}/c^{2}}},\text{ \ }%
x^{\prime }=\frac{x-wt}{\sqrt{1-w^{2}/c^{2}}},\text{ \ }y^{\prime }=y,\text{%
\ \ }z^{\prime }=z,  \label{Lor7}
\end{equation}
interpreted as the transformations of the time and space axes of the two
frames moving in the $x$ direction at relative velocity $w.$ The analogical
interpretation may be given to the formulas 
\begin{equation}
t^{\prime }=t\sqrt{1+v^{2}/c^{2}}-\frac{v}{c^{2}}z,\text{ \ \ }z^{\prime }=z%
\sqrt{1+v^{2}/c^{2}}-vt,\text{ \ \ }y^{\prime }=y,\text{\ \ }z^{\prime }=z,%
\text{\ }  \label{Lor8}
\end{equation}
obtained by means of parametrization (\ref{vw2}). However, the \textit{%
preferred frame description} must refer only to the rest frame of the
observer, what, in general, excludes the interpretation assuming that above
transformation formulas relate, so called, inertial observers. The both
formulas (\ref{Lor7}) and (\ref{Lor8}) concern the same \textit{passive}
transformation (\ref{Lor5}), so that it is better to use the \textit{passive 
}$\eta $ instead of velocity as the transformation parameter. The exception
is the case $\eta \approx 1$ corresponding to small values of $w$ and $v$.
In this situation, the change of \textit{comparison scale }in the \textit{%
preferred frame description }$($given for $\eta =1)$ reduces the formulas (%
\ref{Lor7}) and (\ref{Lor8}) to the Galilean form, and thus provide us the
``true'' classical limit.

\section{\textit{Vital time} as a Lorentz invariant}

Within the framework of the above results it is worthwhile to emphasize once
more the difference between the \textit{frozen} and \textit{vital} time
meanings.

Namely, let us again consider the Lorentz transformations carried out,
however, in a light-cone frame, where are called the \textit{squeeze
transformations} \cite{kim96}. In Minkowski frame the two light-cone axes
are given by the eqs. 
\begin{equation}
x_{+}=ct\text{ \ \ and \ \ }x_{-}=-ct.  \label{lc111}
\end{equation}
The physical interpretation of coordinates $x_{+}$ and $x_{-}$ is clear: in
the upper light cone part ($t\geqslant0$) $x_{+}$ and $x_{-}$ are the
appropriate distances that a light pulse needs to cover in a time period $t$.

Now, let us apply \textit{squeeze transformations} to this frame. The axes
of the new frame $x_{+}^{\prime}$ and $x_{-}^{\prime}$ must satisfy 
\begin{equation}
x_{+}^{\prime}=\frac{1}{\eta}x_{+}\ \ \ \ \text{and \ \ }x_{-}^{\prime}=\eta
x_{-}.  \label{lc112}
\end{equation}
But the same, due to (\ref{lc111}), may be written in an alternative form 
\begin{equation}
t_{+}^{\prime}=\frac{1}{\eta}t\text{ \ \ and \ \ }t_{-}^{\prime}=\eta t,
\label{lc113}
\end{equation}
which interpretation may confuse us a little bit. Indeed, eqs. (\ref{lc113})
cannot be interpret as \textit{vital time} transformations. The vital time
meaning can be assign only to the eq. (\ref{lc111}), whereas the squeeze
transformation (\ref{lc112}), in general, destroy this meaning. Nevertheless
eq. (\ref{lc113}), of course, makes sense but for the \textit{frozen time}
meaning. In other words, any \textit{vital time} interval may be always
identified with the interval of a light cone axis, which equation in a given
Minkowski frame is given by (\ref{lc111}). This means, that the \textit{%
vital time} and so-called proper time is the same.

Having this in mind, let us reexamine the textbook definition of the proper
time, which says that the proper time of a system is the time measured by a
clock which is stick to the system. One shows, that starting from the
interval invariance principle (\ref{Linv5}), the relation between the proper
time $\tau $ and the time $t,$ measured in the frame relative to which the
``proper clock'' moves at the velocity $w,$ is given by 
\begin{equation}
\tau =t\sqrt{1-w^{2}/c^{2}}.  \label{prop}
\end{equation}
One then concludes, that the proper time (of a particle) is always less then
the relevant time in the stationary frame. Within the framework of presented
approach such interpretation, of course, is quite incorrect. This is because
the origin of Einstein interpretation is based on quite classical arguments,
i.e. on prejudice against the quantum ones \cite{epr}, which currently
constitute the foundations for the present analysis.

Indeed, if one put $ct=\lambda_{0},$ (where $\lambda_{0}$ is the Compton
wavelength) than $h/c\tau$ is the corresponding particle energy, i.e. the
energy of a particle, which in a laboratory frame moves at a velocity $w$.
On the other hand, if one looks at the eq. (\ref{prop}) as the relation
between the \textit{vital time} intervals, then its interpretation needs to
be supported by quantum-mechanical analysis given in the preceding section.
One should notice, however, that now $\tau\rightarrow\Delta t$ and $%
t\rightarrow \Delta\tau$ (cf. eq. (\ref{uc11})). Therefore, the both time
intervals indeed have the \textit{vital} meaning, however, this concerns
only the preferred Minkowski-observer frame. Thus, in general, one finds
that the time that undergo relativistic transformation rules is the \textit{%
frozen} one, whereas the \textit{vital }one is the Lorentz invariant.

\section{Concluding remarks}

Excepting the gravity issue, the basic tools of physical description are:
the Newton theory, quantum mechanics and special relativity. Although these
three theories have been developed as quite independent ones, currently they
provide the basis of the quantum field theory. Of course, it is commonly
thought that the known Dirac equation is the element that joins the two
separate realms of relativity and quantum mechanics. Nevertheless, the
interpretation of the Lorentz symmetry presents simply a generalization of
the interpretation of the Galilean transformation in the Newton theory, what
finds its reflection in the first Einstein relativity postulate. The main
purpose of this paper was to suggest that the origin of the Lorentz symmetry
is rather quantum then classical.

This outlook results from given elementary analysis embracing the idea of 
\textit{preferred frame description} (and observation),\ the concept of 
\textit{relative scaling,} and the \textit{light-cone skeleton} able to
encompass basic informations about external and internal particle features.
In our analysis these external features were particle energy and momentum,
whereas the internal ones were particle space-time extensions.

It has been shown that the \textit{relative scaling }combined with the
principles of quantum mechanics yield clear interpretation of Lorentz
transformations, devoid of a tinge of guessing. Let us note, that the
interpretation of the Lorentz symmetry as the consequence of the relativity
principle had come as a result of the replacement of the Galilean
transformation by the Lorentz one, in a situation when mathematical
structure of the Maxwell equations had made impossible to apply the Galilean
transformation. However, there had been significant differences in
foundations of both (i.e. Newton and Maxwell) theories, which had caused
that such generalization might be not justifiable. For example, in the
Newton theory we deal with the concept of material point, which does not
have a simple counterpart in the wave-like description. Although the Maxwell
equations are seen as the classical ones (and the Plank constant cancels out
from the photon dispersion relation), the simple physical observations, such
as the photoelectric effect, or the blackbody radiation, revel the quantum
structure of the electromagnetic field. Thus, one may say that the
electromagnetic field is a system in which the classical and quantum
features are somehow ``smeared out''. Nevertheless, the Lorentz symmetry so
far is always thought as a manifestation of the classical (space-time)
property. The analysis given in the paper suggests, that just the quantum
features of physical systems decide that Lorentz or Poincar\'{e} symmetry
comes into play.

Briefly, there are two separate ways of approaching the matter of
relativistic quantum physics. The first, well-known, assumes relativity of
frame description, relativity of time and existence of material point-like
objects. The other, currently proposed, assumes preferred frame description,
absolute time and extended quantum objects with internal structures. One of
the consequences of the latter approach is that special relativity, in a
natural way, becomes an integral part of quantum mechanics.

\bigskip

\textit{Acknowledgment}: the author of the paper is very grateful to
Professor Y. S. Kim, as well as, to Professors S. Duplij, W. Siegel and J.
Bagger, the Editors of Kluwer SUSY Encyclopedia, for received suggestions
helpful in preparation of this article.

\end{document}